\documentclass[letterpaper,11pt]{article}
\usepackage{amsmath, amsthm, amssymb}    % May not all be necessary
\usepackage{bridges}			% Custom bridges proceedings style
\usepackage{graphicx}			% For including pictures
\usepackage[colorlinks=true, urlcolor=blue, citecolor=black, linkcolor=black]{hyperref}  
\usepackage{subcaption}			% For captioning multi-panel figures
\usepackage{listings}
\usepackage{xcolor}

\newcommand{\cindyjs}{\emph{CindyJS}}

\newcommand{\cindyscript}{\emph{CindyScript}}
\newcommand{\nyka}{\emph{Nyka}}
\newcommand{\init}{\emph{init}}
\newcommand{\draw}{\emph{draw}}
\newcommand{\tick}{\emph{tick}}
\newcommand{\mousedown}{\emph{mousedown}}

\definecolor{backcolor}{rgb}{0.95,0.95,0.96}
\definecolor{commentcolor}{rgb}{0.5,0.5,0.5}
\definecolor{keywordcolor}{rgb}{0.2,0.3,0.8}
\definecolor{stringcolor}{rgb}{0.1,0.5,0.1}
\definecolor{numbercolor}{rgb}{0.2,0.3,0.8}

\title{From Code to Canvas}

\author{Bernhard O. Werner\textsuperscript{1}\vspace{10pt}\\
\textsuperscript{1}Munich University of Applied Sciences; bernhard.werner@hm.edu} % end \author
\date{}					% Suppress any date on submissions

\begin{document}
\maketitle

% Prevent page number 1 from being printed on the first page.
\thispagestyle{empty}

\begin{abstract}

The web-based dynamic geometry software \cindyjs\ is a versatile tool to create interactive applications for mathematics and other topics. In this workshop, we will look at a code package that makes the creation of animations in \cindyjs\ easier and more streamlined. Animations, which can then be embedded into presentations or be used in (lecture) videos. The focus lies on the creation of the animations themselves and some of the technical and artistic fundamentals to do so. 
% We will only touch on exporting and editing, and entirely ignore how to incorporate the whole process into the production pipeline of, say, a YouTube channel or lecture series.
\end{abstract}

%%%%%%%%%%%%%%%%%%%%%%%%%%%%%%%%%%%%%%%%%%%

\section*{CindyJS}
\label{sec:cindyjs}
% \cinderella\ is a dynamic geometry software that was originally developed by Jürgen Richter-Gebert and Ulrich Kortenkamp in the late 1990s. It allows users to create and manipulate geometric constructions in an interactive and visual way. Users can export constructions as web applications, allowing for broader sharing and interactive demonstrations online. For the last decade, this has been done via \cindyjs: a partner project, that runs in the browser using JavaScript. This eliminates the need for any installations by end users, making it handy to distribute the applets created with it. Except for minor differences, it is fully compatible with \cinderella . These differences are constantly being ironed out, though.

\cindyjs\ is a dynamic geometry software that was originally developed by Jürgen Richter-Gebert and Ulrich Kortenkamp. It allows users to create and manipulate geometric constructions in an interactive and visual way. Cf. \cite{G:cindyjs}. It runs in the browser using \emph{JavaScript}. This eliminates the need for any installations by end users, making it handy to distribute the applets created with it.

After many years of building interactive applications with \cindyjs , I developed a package for it to streamline the process of creating programmatic animations. This package is the topic of this paper and the associated workshop.

For the following discussions, I will assume that you are familiar with the fundamentals of programming itself. If you want to learn the scripting language \cindyscript\ properly, I recommend the book \cite{RGK:cindy_manual}, looking at the documentation and the example gallery on \url{https://cindyjs.org}, as well as the examples at \cite{R:cindyjs}. It should still be possible to follow along and at least read the code without much programming experience. In addition, you can find a cheat sheet under \url{https://bit.ly/cheat25}, which explains the main points to pay attention to in \cindyjs\ and the core features of the animation package.

\section*{The Setup}
\label{sec:setup}

We will use an online editor that lets us jump right into coding animations. You can find it via the URL:
\begin{center}
    \url{https://bit.ly/bridges25}
\end{center}

\begin{figure}[h!tbp]
	\centering
	\includegraphics[width=6.5in]{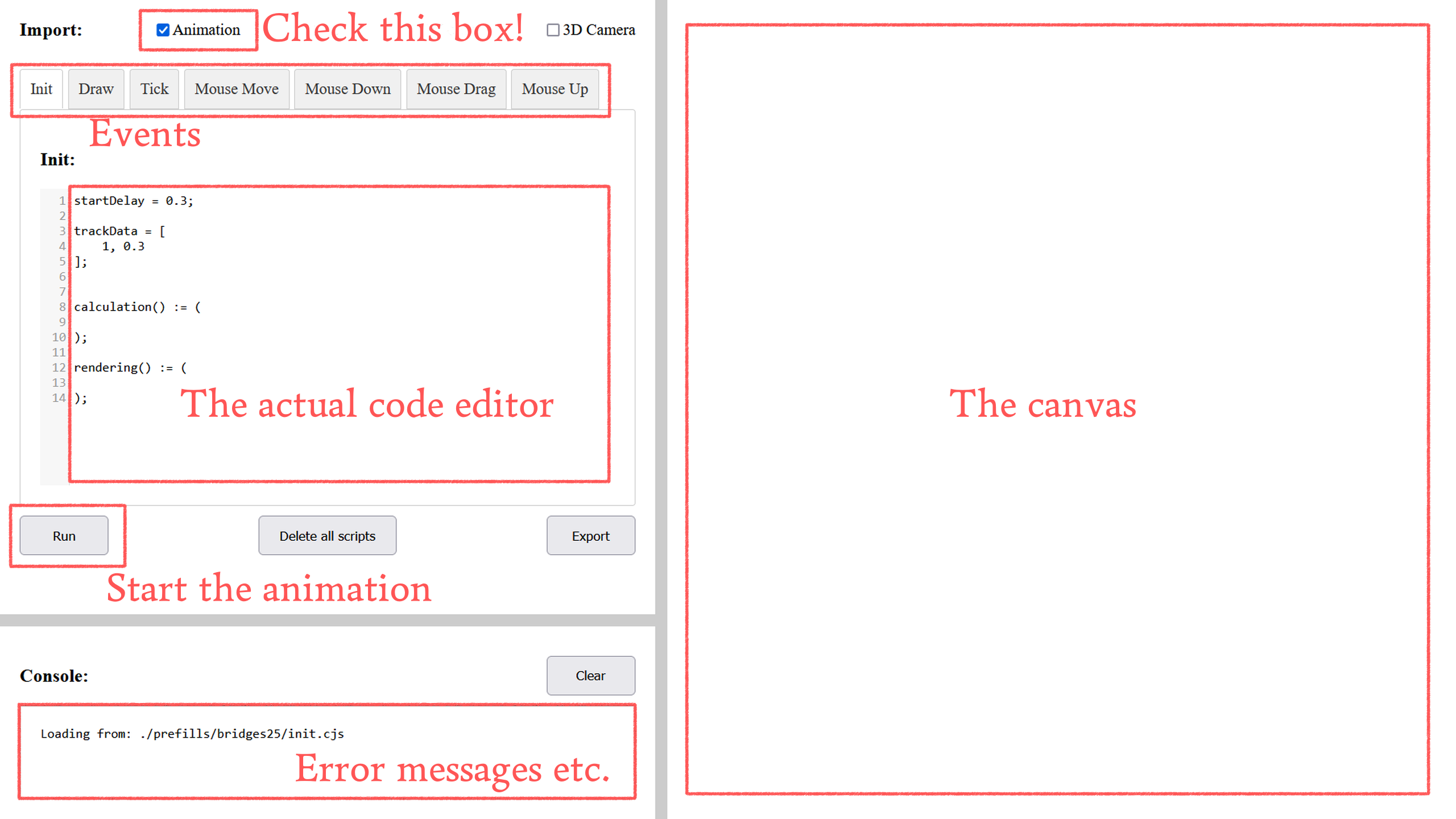}
	\caption{The web editor found at \url{https://bit.ly/bridges25} with its main features marked.}
	\label{fig:editor}
\end{figure}

You should see something similar to Figure \ref{fig:editor}. Note that this editor is a work-in-progress; some details might change between me writing this and you reading it. Moreover, if there are some glitches, please refresh the page. The key parts of the editor are the following: 

We have the canvas we will draw on on the right-hand side. Plus the console in the bottom-left of the screen which shows us error messages. In the top-left we have the code editor where the actual work is done. It starts with checkboxes for various packages, of which only \emph{animation} is important for us here. Below, we can select different so-called events we can attach code to. The \init\ event, for example, is triggered when the program starts; \draw\ when the objects that are actually drawn on screen are updated; and \mousedown\ when a mouse button is pressed; etc. The animation package refactors a few things so that we can do all our work in the \init\ event without having to worry about the others. Next, we get some buttons, of which "Run" is the important one. It starts the animation; but that can also be done with the keyboard command \emph{Ctrl~+~Enter}. In the actual editor section, you can see some boilerplate code that is already pasted in and which should be similar to this here:
\begin{lstlisting}
startDelay = 0.3;

trackData = [
    1, 0.3,
    2, 0.5
];

calculation() := (

);

rendering() := (

);
\end{lstlisting}

If you press "Run", you should see some debugging information in the canvas and, in particular, the total time ticking away. The structure of this code scaffold is the following: In lines 3 to 6, so-called animation tracks are set up. They are the way this package keeps track of the progress of each animation in the sequence. A track itself is specified by its duration and the length of the pause to the next animation, which alternate in the array \lstinline{trackData}. Here, we have two placeholder tracks. The first lasts 1s and has a 0.3s pause after it. The second takes 2s with a 0.5s pause afterwards. For more tracks, we simply extend the list; cf. Example 2. Note that you always have to specify a pause after the last track. The optional variable \lstinline{startDelay} in line 1 determines how long to wait with starting the animations after the website loads.

The package also creates progress variables for all tracks that run from 0 to 1. We will use these to link the time to the properties we want to animate. These progress variables always start with a \lstinline{t} followed by the number of the track it belongs to: \lstinline{t1}, \lstinline{t2}, \lstinline{t3}, etc. On lines 8 to 14, we have placeholders for two functions that do the actual animating: \lstinline{calculation} which is later on called within the \tick\ event and \lstinline{rendering} which is called in the \draw\ event. The \tick\ event runs (ideally) 60 times per second and is the \cindyjs\ event that progresses the internal clock of the package and updates the animation tracks accordingly. And \draw , as was already mentioned, actually draws things on screen. The basic idea here is to do all the (complicated) calculations that are necessary to update various objects---like changes in position, size or colour---in \lstinline{calculation}. All draw commands go into \lstinline{rendering}.

\begin{figure}[h!tbp]
	\centering
	\includegraphics[width=6.5in]{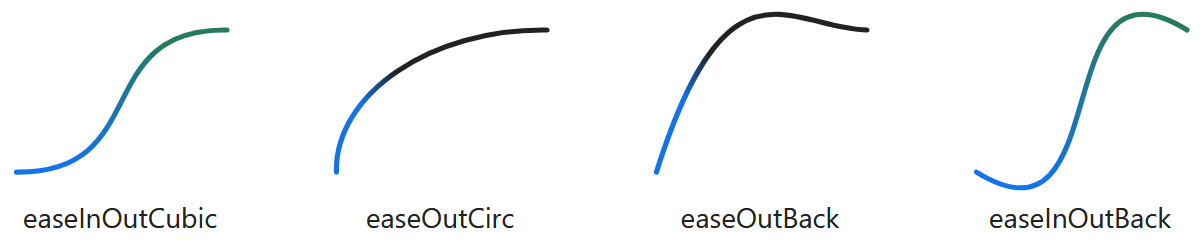}
	\caption{The four easing functions used in this paper. Screenshots taken from \cite{SS:easing}.}
	\label{fig:easing}
\end{figure}

\section*{Examples}

The following examples illustrate some of the core concepts of the animation package. More can be found at \cite{W:cindy_tools}. If you are working through the examples on your own, you should type the proposed code into the code editor linked above and, much more importantly, try to change and expand the code as much as possible. E.g. in the first example, we will create a single straight line, and you should immediately try to make a simple coordinate grid. This will greatly foster your understanding of how this package works.

\subsection*{Example 1: a Line}

To start, let us animate a line segment being drawn between two points. After specifying these points, the animation itself will be achieved by calculating a point that moves from the start to the end point. This is done via a function called \lstinline{lerp} which is short for “linear interpolation”. It is defined as 
\begin{lstlisting}[numbers=none]
lerp(x, y, t) := t * y + (1 - t) * x;
\end{lstlisting}

\begin{figure}[h!tbp]
	\centering
	\includegraphics[width=6in]{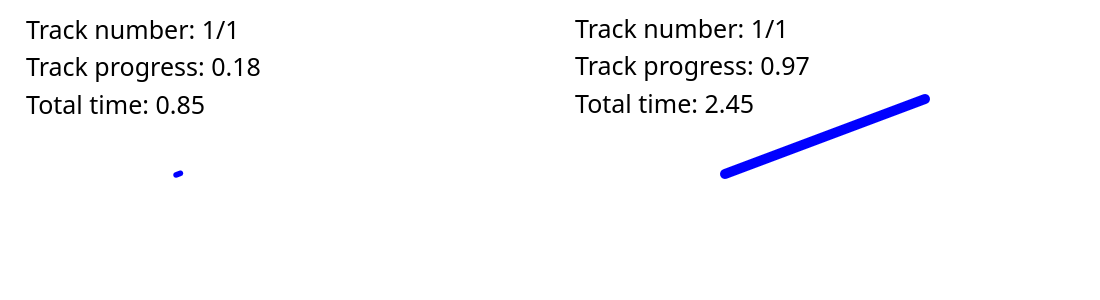}
	\caption{The line from Example 1 at two points in time of the animation.}
	\label{fig:exa_line}
\end{figure}

Most things you see on a computer screen are drawn with a version of this function. The idea is that the value \lstinline{t} represents how close the result is to \lstinline{x} and \lstinline{y}. If \lstinline{t} is exactly 0, we are at \lstinline{x}. If it is 1, we are at \lstinline{y}. And if it is 0.5, we are right in the middle between \lstinline{x} and \lstinline{y}. In \cindyjs , this works for all objects for which the right-hand side can be evaluated. So, not only for numbers, but also for (nested) arrays of equal size which are interpreted as vectors or matrices. This transition from one object to another is then driven by the progress variable of an animation track. That means, after defining the coordinates of the two points \lstinline{a} and \lstinline{b} of our animation, we could calculate a moving end point between them as: \lstinline|endPoint = lerp(a, b, t1);|

Using the progress of the animation track directly, however, would lead to a stiff looking animation. It is common to use so-called easing functions to give the movement more character. The most common one is a cubic polynomial that starts and ends slowly---evoking the sense of physical acceleration. In the package, it is called \lstinline{easeInOutCubic}. The package also provides all other easing functions found in \cite{SS:easing}. The easing functions used in this paper can be seen in Figure \ref{fig:easing}. Adding all of that to the boilerplate code in the online editor could lead to something like this:

\begin{lstlisting}
startDelay = 0.5;
trackData = [
  2, 0.5
];

a = [-3, 7];
b = [5, 10];

calculation() := (
  endPoint = lerp(a, b, easeInOutCubic(t1));
);

rendering() := (
  draw(a, endPoint, size -> 10 * easeOutCirc(t1));
);
\end{lstlisting}

In line 3, a single 2-seconds long animation track is defined. Lines 6 and 7 specify the coordinates of the actual end points we want to draw a segment between. The moving end point calculation is added in line 10 inside the \lstinline{calculation} function. And the actual drawing of the segment is done inside the \lstinline{rendering} function on line 14 by connecting the start point \lstinline{a} with the moving \lstinline|endPoint|. Note that we also modify the thickness of the line segment over time here. Which is one of the many ways to customize the style of an animation.

\begin{figure}[h!tbp]
	\centering
	\includegraphics[width=6in]{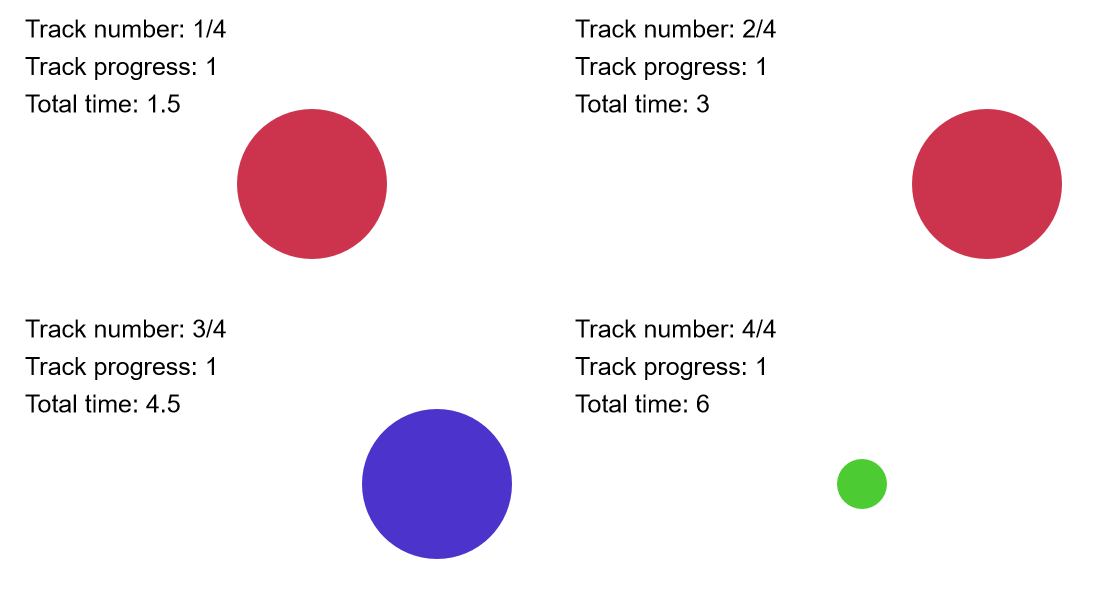}
	\caption{The circle from Example 2 at the end of each animation track.}
	\label{fig:exa_circle}
\end{figure}

\subsection*{Example 2: a Circle}
Usually, you will want to have a sequence of animations, for which we need multiple animation tracks. Say, we want to make a circle appear. Next, it should move to the right. Then, it should change its colour. And finally, it should move back left, change colour again and shrink a bit---all at once. By and large, this works exactly like Example 1, just with more steps: We use a couple of \lstinline{lerp} commands to change the variables that determine the circle. The difference is that we define them inside of the \lstinline{calculation} function, since we want to change the values itself instead of calculating a new one as in the previous example. This allows us to write something like
\begin{lstlisting}[numbers=none]
property = lerp(property, newValue, t);
\end{lstlisting}
in order to move a value from whatever it previously was to a new value. With this change in mind, we get the following code for the animation described above:

%In contrast, in Example 1, we calculated a completely new value based on the points \lstinline{a} and \lstinline{b} because of which we were able to define them before the \lstinline{calcluation} function. 

\begin{lstlisting} 
startDelay = 0.5;
trackData = [
  1, 0.5,  //  01    create circle 
  1, 0.5,  //  02    move circle
  1, 0.5,  //  03    recolor circle
  1, 0.5   //  04    move, scale and recolor circle
];

calculation() := (
  // Define circle properties.
  circleCenter = canvasCenter;
  circleRadius = 0;
  circleColor = (0.8, 0.2, 0.3);  

  // Animate circle properies.
  circleRadius = lerp(circleRadius, 3, easeOutBack(t1));
  
  circleCenter = lerp(circleCenter, canvasCenter + [10,0], easeInOutCubic(t2));
  
  circleColor = lerp(circleColor, (0.3, 0.2, 0.8), t3);
  
  circleRadius = lerp(circleRadius, 1, easeInOutBack(t4));
  circleCenter = lerp(circleCenter, canvasCenter, easeInOutCubic(t4));
  circleColor = lerp(circleColor, (0.3, 0.8, 0.2), t4);
);

rendering() := (
  fillcircle(circleCenter, circleRadius, color -> circleColor);
);
\end{lstlisting}

Note that on the lines 3 to 6, where the durations and pauses of the animation tracks are set up, we have comments at the end of each line describing what the respective track is doing. This is a handy planning process for longer animation sequences: Start by writing these comments to outline the animation, and specify the durations of and pauses after each track. Only then write the code for the actual animation.

\subsection*{Example 3: a Curved Arrow}
Another common object to draw for a maths animation is a curved line or, in this example, a curved arrow. The animation package achieves this by sampling points on certain classes of curves. We can then gradually draw more and more line segments between consecutive points. If the points are close enough, it will look like the whole curve is drawn smoothly. One of the available classes of curves are Bézier curves. They are a staple in graphics design and animation and the go-to method to model curved lines. They are determined by a series of control points, of which the first and last represent the start and end point of the curve. The control points in-between dictate the shape. See \cite{K:bezier} for more details---but the general idea is that the curve will move towards each control point, but not reach it. In code, we would write something like this:
\begin{lstlisting}[numbers=none]
stroke = sampleBezierCurve([(8, 12), (0, 15), (10, 4), (-2, 4)], resolution);
\end{lstlisting}
The first argument of the function \lstinline{sampleBezierCurve} is the list of control points. The second argument determines the sample rate. The higher it is, the smoother the resulting animation. However, the frame rate might drop if you put it in \lstinline|calculation|. And the result is a list of \lstinline{resolution}-many points along the curve. After this pre-calculation, we can render the stroke like this:
\begin{lstlisting}[numbers=none]
strokeIndex = round(lerp(1, resolution, easeInOutCubic(t1)));
connect(stroke_(1..strokeIndex));
\end{lstlisting}

\begin{figure}[h!tbp]
	\centering
	\includegraphics[width=6in]{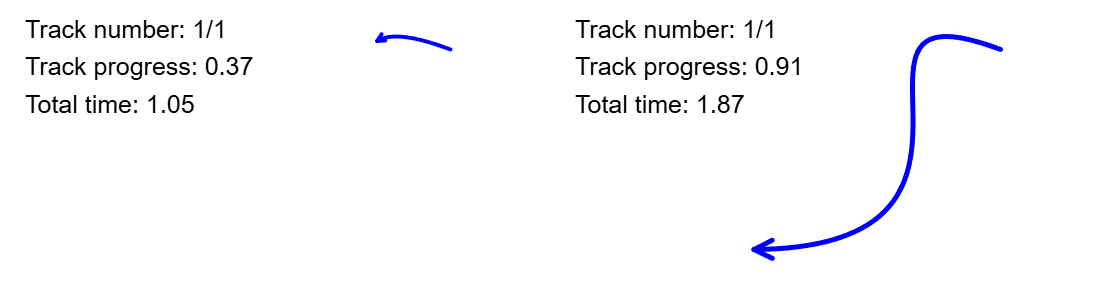}
	\caption{The arrow from Example 3 at two points in time of the animation.}
	\label{fig:exa_arrow}
\end{figure}

The function \lstinline{connect} takes a list of points and draws line segments between consecutive points; exactly what we want here. We get the desired animation since more and more points are taken into account over time in the code above. This would draw the curve itself, but as mentioned at the start, we want an arrow. For that, we draw the very last segment of the stroke separately using the arrow modifier. E.g.,
\begin{lstlisting}[numbers=none]
draw(stroke_(strokeIndex - 1), stroke_strokeIndex, size -> 5 * easeOutCirc(t1),
    arrow -> true, arrowsize -> 2 * t1);
\end{lstlisting}

But! This, of course, only works if we have at least two points to connect. So, we need an if-statement to check for that. Taking all of that into account, the code for the full animation of a curved arrow might look like this:
\begin{lstlisting} 
startDelay = 0.5;
trackData = [
  1.5, 0.5
];

resolution = 256;
stroke = sampleBezierCurve([(8, 12), (0, 15), (10, 4), (-2, 4)], resolution);

calculation() := (
    strokeIndex = round(lerp(1, resolution, easeInOutCubic(t1)));
);

rendering() := (
  connect(stroke_(1..strokeIndex), size -> 5 * easeOutCirc(t1));
  if(strokeIndex >= 2,
    draw(stroke_(strokeIndex - 1), stroke_strokeIndex, size -> 5 * easeOutCirc(t1), arrow -> true, arrowsize -> 2 * t1);
  );
);
\end{lstlisting}

\subsection*{Example 4: Text}
The last common element of a maths animation are text and maths formulas. Drawing text in \cindyjs\ works by specifying the position and the text string itself, plus typical modifiers like size, colour, or font family:
\begin{lstlisting}[numbers=none]
drawtext([0,0], "lorem ipsum", size -> 30, color -> (1, 0, 0.5), family -> "Arial");
\end{lstlisting}
This also works with TeX formulas. The command 
\begin{lstlisting}[numbers=none]
drawtext([0,0], "$\sum_{i=0}^\infty q^i= \frac{1}{1-q}$");
\end{lstlisting}
would render $\sum_{i=0}^\infty q^i= \frac{1}{1-q}$ on the canvas. You can lerp the position and opacity (via the modifier \lstinline|alpha|) to create slide-in and fade-in animations for text. If you want something more dynamic like a typewriter-like effect where the characters appear one after another, we need a bit more work: To achieve something like that, the package separates a string into individual glyphs via a function called \lstinline{fragment}. It takes the string itself as an argument, together with the desired text size. Now, for many reasons, I decided to create a custom typesetting language called \nyka\ for this process. By and large, it is identical to \emph{LaTeX}, but the syntax is sometimes slightly different. For example, in \nyka\ you would write
\begin{lstlisting}[numbers=none]
"$\sum[i=0][\infty] q^i$"
\end{lstlisting}
instead of
\begin{lstlisting}[numbers=none]
"$\sum_{i=0}^\infty q^i$"
\end{lstlisting}
A full list of commands in which \nyka\ differs from \emph{LaTeX} can be found at \cite{W:cindy_tools}. To use this for animations, there is one last catch: \cindyjs\ uses a plugin that's based on an old release of \textit{KaTeX} (cf. \cite{EA:katex}). It has to load the fonts at the very start. But that happens after the \init\ script is executed. Everything that uses the exact font information must therefore be placed inside a function called \lstinline{delayedSetup} which handles this exact case. That is true for \lstinline{fragment} which needs size information of the glyphs in the text to position them correctly. That means that the way to animate a typing-effect for text looks like this:

\begin{lstlisting}
textSize = 30;
typingSpeed = 10;

string = "This is $\displaystyle \sum[i=0][\infty]q^i = \frac{1}{1-q}$ plus text.";
fragmentedString = fragment(string, textSize);

startDelay = 0.5;
trackData = [
  fragmentLength(fragmentedString) / typingSpeed, 0.3
];
    
delayedSetup() := (
  fragmentedString = fragment(string, textSize);
);

rendering() := (
  drawFragments([-5, 12], fragmentedString, t1, "up", {"color": (0.8, 0.2, 0.3)});  
);
\end{lstlisting}

In line 5, we call \lstinline{fragment} once. At this point, the \emph{KaTeX} fonts aren't loaded yet. But it will correctly calculate the number of glyphs in the text. This is then used in line 9 with the help of the function \lstinline{fragmentLength} to determine the duration of the animation track. Usually, it looks best if all text throughout an animation sequence is typed at the same speed. Specifying this speed and calculating the duration of the animation like here is the simplest way to achieve this. In line 13 we then call \lstinline{fragment} again within \lstinline{setupAfterKatex}, as explained above. Finally, the actual drawing is done via the function \lstinline{drawFragments} in line 17. Its first two arguments are the position and the deconstructed text. Next comes the progress variable of the desired animation track. The fourth argument is the typing mode. Here, \lstinline{"up"} means that the characters appear from below and move upwards to their position. The last arguments are the usual modifiers. Alas, modifiers only work properly for built-in functions. As of writing this paper, it's better to hand them over in a dictionary like this and process them yourself in custom functions.

\begin{figure}[h!tbp]
	\centering
	\includegraphics[width=6in]{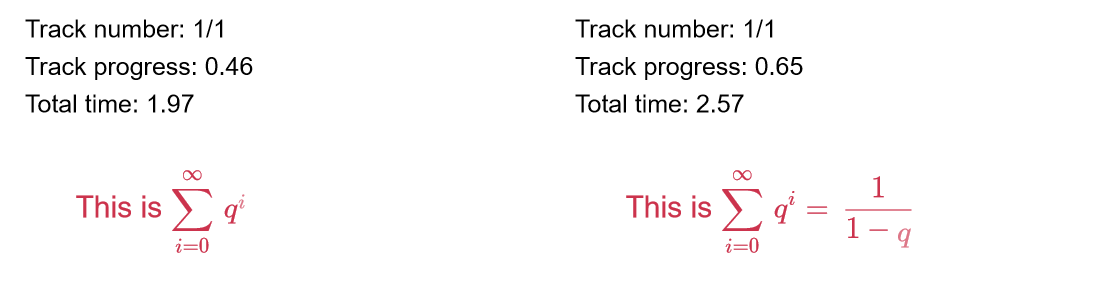}
	\caption{The text from Example 4 at two points in time of the animation.}
	\label{fig:exa_text}
\end{figure}

\section*{Everything That Wasn't Covered}
\label{sec:outlook}
There is obviously much more to say about creating (good) animations with this package. First of all, the documentation and examples under \cite{W:cindy_tools} provide a fuller picture of how typical animation sequences can be achieved. Another big topic is how to actually turn an animation into a video. Again, \cite{W:cindy_tools} offers more details, but the gist is: You add the lines
\begin{lstlisting}[numbers=none]
showDebugInfo = false;
renderMode = RENDERMODE.FRAMES;
\end{lstlisting}
to the top of the \init\ script. The first one turns off the debugging information in the top-left corner. The second one will tell the package to render the animation frame by frame at a fixed 60 FPS. You can double-click on the canvas and the package will start generating all frames of the animation and download them to your computer as sequentially named PNGs. Then, you can use a program like \emph{FFmpeg} or \emph{DaVinci Resolve} to turn them into a video.

In the workshop itself, we can hopefully go through more examples and background information. This paper, however, should have covered the basics to get everyone started with programmatic animations using this CindyJS package.

{\setlength{\baselineskip}{13pt} % tighten line spacing for bibliography
\raggedright				% no right justification for References

} % end setlength, raggedright
   

\begin{thebibliography}{99}
\bibitem{EA:katex} E. Eisenberg \& S. Alpert. \textit{KaTeX}, 2025. URL: \url{https://katex.org/}

\bibitem{G:cindyjs} M. von Gagern, U. Kortenkamp, J. Richter-Gebert, \& M. Strobel. "CindyJS. Mathematical visualization on modern devices." \textit{Mathematical Software – ICMS 2016: 5th International Conference}, Springer, 2016, pp. 319–326. \url{https://doi.org/10.1007/978-3-319-42432-3_39}

\bibitem{K:bezier} M. Kamermans. \textit{A Primer on Bézier Curves}, 2020. URL: \url{https://pomax.github.io/bezierinfo/}

\bibitem{R:cindyjs} J. Richter-Gebert et al. \textit{CindyJS} [Source Code], 2025. URL: \url{https://github.com/CindyJS/CindyJS}

\bibitem{RGK:cindy_manual} J. Richter-Gebert \& U. H. Kortenkamp. \textit{The Cinderella. 2 manual: working with the interactive geometry software.} Springer Science \& Business Media, 2012.

\bibitem{SS:easing} A. Sitnik \& I. Solovev. \textit{Easing Functions Cheet Sheet}, 2024. URL: \url{https://easings.net/}

\bibitem{W:cindy_tools} B. O. Werner. \textit{Cindy Tools} [Source Code], 2025. URL: \url{https://github.com/BernhardWerner/cindy_tools}

\end{thebibliography}
\end{document}